%% file: main.tex
\documentclass{article}

\pdfoutput=1

\usepackage{newtxmath} 
\usepackage{newtxtext} 
\usepackage{fullpage}

\usepackage{authblk}
\usepackage{natbib}
\usepackage[utf8]{inputenc}
\usepackage{amsmath}
\usepackage{subcaption}
\usepackage{xcolor}
\usepackage{trimclip}
\usepackage{graphicx}
\usepackage{listings}

\title{Estimating volcanic ash emissions using retrieved satellite ash columns and inverse ash transport modelling}

\author[1]{André R. Brodtkorb}
\author[1]{Anna Benedictow}
\author[1]{Heiko Klein}
\author[2]{Arve Kylling}
\author[1]{Agnes Nyiri}
\author[1]{Alvaro Valdebenito}
\author[2]{Espen Sollum}

\affil[1]{The Norwegian Meteorological Institute, Oslo, Norway}
\affil[2]{NILU - Norwegian Institute for Air Research, Kjeller, Norway}

\renewcommand\Affilfont{\itshape\small}

\newcommand{\emist}{\alpha}
\newcommand{\emisa}{\beta}
\newcommand{\obslat}{y}
\newcommand{\obslon}{x}
\newcommand{\obstime}{t}

\graphicspath{{svg-inkscape/}}

\begin{document}

\maketitle

\begin{abstract}
    This paper describes the inversion procedure being used operationally at the Norwegian Meteorological Institute for estimating ash emission rates from retrieved satellite ash column amounts and a priori knowledge. 
    The overall procedure consists of five stages: 
    (1) generate a priori emission estimates;
    (2) run forward simulations with unit emissions; 
    (3) collocate/match observations with emission simulations;
    (4) build system of linear equations; and
    (5) solve overdetermined system.
    We go through the mathematical foundations for the inversion procedure, performance for synthetic cases, and performance for real-world cases. The novelties of this paper includes pruning of the linear system of equations used in the inversion and inclusion of observations of ash cloud top altitude.
    %
    The source code used in this work is freely available under an open source license, and is possible to use for other similar applications.  
\end{abstract}

\section{Introduction}
Determining ash emissions during a volcanic eruption is important in order to be able to give realistic forecasts of volcanic ash transport in the atmosphere. There are typically several types of observations of the eruption, such as ground sightings, plane/helicopter sightings, and satellite images. The sightings may give observations of ash cloud height, and satellite images may give spatial extent as well as ash mass estimates of the vertical air column away from the volcano. The plume height observations can be used to estimate the mass eruption rate using e.g. formula from \cite{mastin2009multidisciplinary}, however, this approach may not provide a realistic estimate of the vertical distribution of the ash. The ash inversion procedure attempts to remedy this by using satellite ash column densities and inverse transport modelling to give a better ash emission estimate.

The main idea of the ash emission estimation procedure presented here is based upon the variational principle, i.e., to use a set of forward simulations with unit emissions, and then attempt to find the linear combination of these that best matches the observed ash locations and concentrations. The methodology used here has earlier been used for sulphur dioxide emission, see e.g.,~\cite{seibert2000inverse, seibert2011uncertainties} and \cite{eckhardt2008estimation}. It was for the first time used for volcanic ash emission rate determination for the Eyjafjallaj{\"o}kull 2011 eruption \citep{stohl2011determination}. \citet{steensen2017uncertainty} presented an uncertainty assessment of the method. This work extends the approach to also incorporate ash cloud height as an observation in the inversion. The manuscript is organised as follows: In Section~\ref{sec:matform} the theory behind the inversion procedure is outlined. The atmospheric dispersion model is described in section~\ref{sec:adm}.The synthethic benchmark cases and real-world cases are described in sections~\ref{sec:synth} and \ref{sec:realworld}, respectively. The paper is summarized in section~\ref{sec:summary}.

\section{Mathematical formulation}
\label{sec:matform}
We base our inversion procedure on the approach taken by \citet{seibert2000inverse}, and adopted to ash emission by several authors (see e.g., \cite{stohl2011determination} and the references therein). 
The inversion procedure is based upon creating a so-called source-receptor matrix, $M$, from a set of forward simulations with unit emissions, 
\begin{align}
    \{S\} = \{ S(\emist_1, \emisa_1), \dots, S(\emist_1, \emisa_l), S(\emist_2, \emisa_1), \dots S(\emist_k, \emisa_l) \} = \{S_1, \dots, S_{n}\}.
\end{align}
Here, $\emist_k$ denotes \emph{emission time} $k$ and $\emisa_l$ denotes \emph{emission level} $l$. An individual simulation $S_{j}$ then contains time-dependent three-dimensional simulation results of a unit of ash emitted into the atmosphere at the given emission time and emission level.

We equivalently have the set of observations,
\begin{align}
\nonumber
    \{O\} &= \{O(\obslat_1, \obslon_1, \obstime_1), O(\obslat_2, \obslon_2, \obstime_2) \dots O(\obslat_m, \obslon_m, \obstime_m)\} \\
    &= \{O_1, \dots, O_m\},
\end{align}
in which $(x_i, y_i)$ denotes the spatial coordinates, $(t_i)$ the time, and the observations are sorted according to increasing time, $\obstime_i$. We can assemble the matrix $M$ so that element $(i, j)$ of the matrix has simulation results from emission $j$ at observation coordinate $i$,
\begin{align}
    M_{i, j} := S_j(\obslat_i, \obslon_i, \obstime_i),
\end{align}
and similarly for the vector of observations so that element $i$ corresponds to observation coordinate $i$,
\begin{align}
	{y_0}_{i} := O_i.
\end{align}

\begin{figure*}
    \centering
    \begin{subfigure}[c]{0.37\linewidth}
    \centering
    %
    \resizebox{\linewidth}{!}{%
            \clipbox{5.5cm 4cm 10.5cm 2cm}{%
                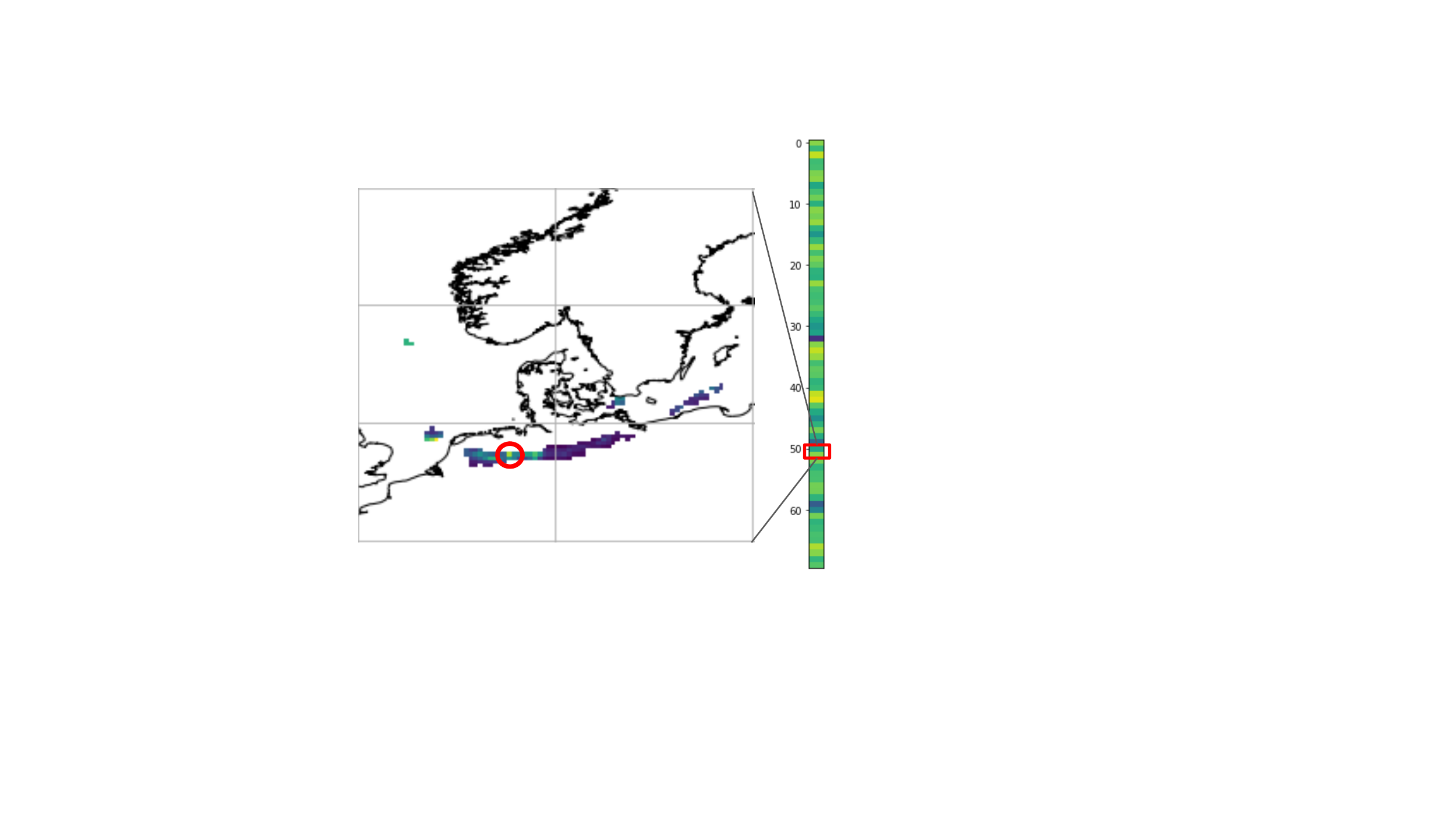%
            }%
    }

    \caption{}
    \label{fig:observations_vector}
    \end{subfigure}%
    \hfill%
    \begin{subfigure}[c]{0.6\linewidth}
    \centering
    %
    \resizebox{\linewidth}{!}{%
            \clipbox{3cm 0.5cm 3.5cm 0cm}{%
                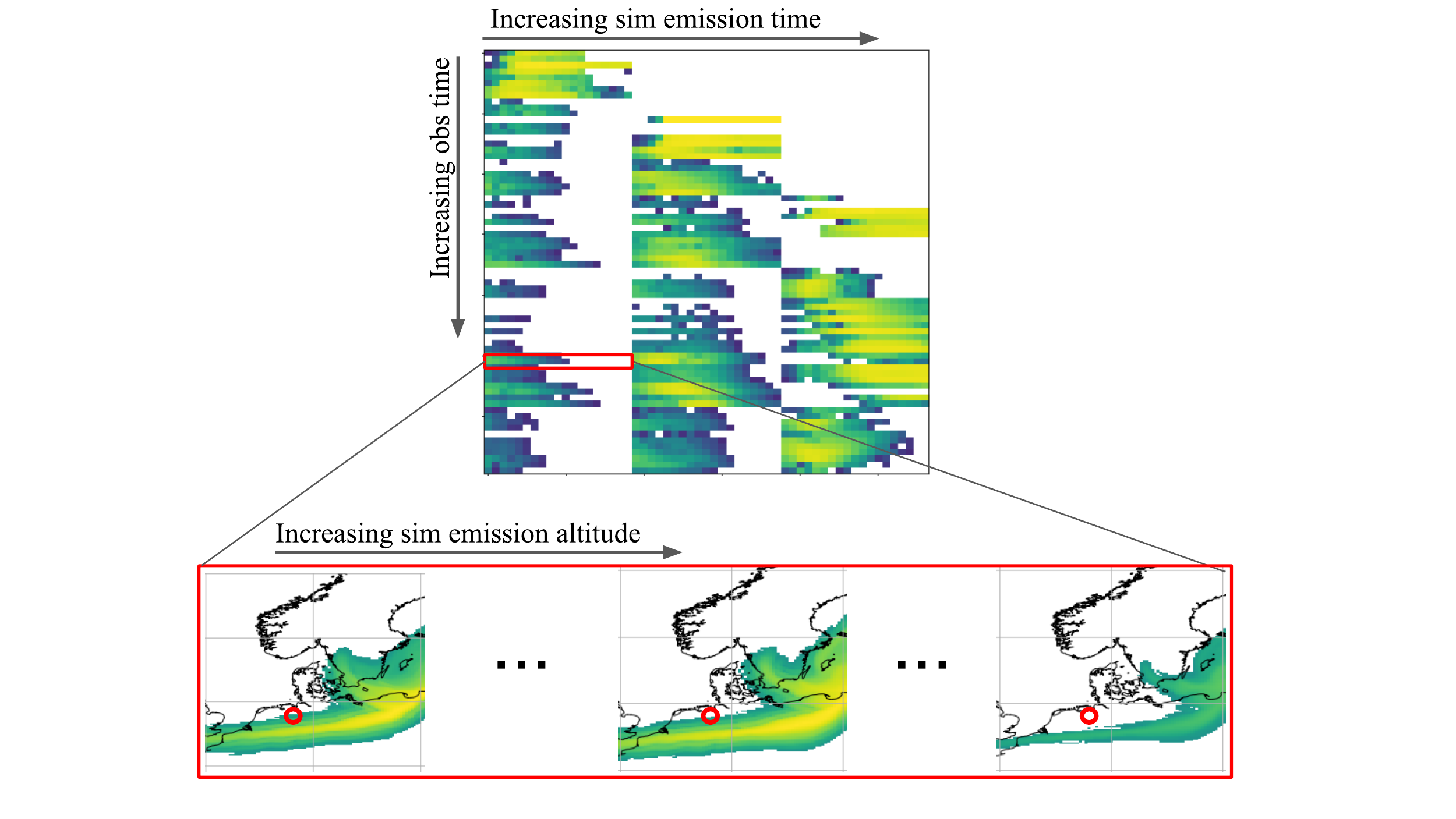%
            }%
    }

    \caption{}
    \label{fig:source_receptor_matrix}
    \end{subfigure}
    \caption{Linear system of equations. %
    (\subref{fig:observations_vector}) shows the vector of observations, in which a single location is highlighted with its corresponding location in the observations vector. 
    (\subref{fig:source_receptor_matrix}) shows the source-receptor matrix $M$. Each row in the matrix corresponds to one observation of ash, $O_j$, and each column corresponds to one emission simulation $S_j$. This matrix shows 60 observations (columns), three emission time points, and 19 emission altitudes ($3\times 19$ rows).
    White means no ash in the simulation, and the colored elements correspond to the concentration of ash in the individual simulations. 
    }
    \label{fig:linear_system}
\end{figure*}

Figure~\ref{fig:linear_system} shows a small source-receptor matrix, and part of the inputs used to create it. For each observation at $(\obslat_i, \obslon_i, \obstime_i)$, we find the simulated ash content at the same time and spatial coordinate for all of the different emission simulations. 

The aim of the inversion procedure is to use this source-receptor matrix, and vector of observation to find the vector $x$ so that 
\begin{align}
Mx = y_0.
\label{eqn:linear_system}
\end{align}
Here, $x$ is the linear combination of unit emissions that best reproduce the observations.

The size of the matrices and vectors involved in the computation is determined by the number of observations and a priori emission values. The number of a priori simulations/emission estimates, $n$, can typically be a few hundred to a few thousand, and the number of observations, $m$ can be hundreds of thousands. For example, for the
2010 Eyjafjallajökull eruption, we have an emission simulation starting every three hours from 00:00 on April 14th UTC to 00:00 on April 18th, and emission heights every 650 meters from 325~m to about 12~km. This corresponds to $33$ distinct times and $19$ elevations, totaling to $627$ unique a priori emission estimates with one simulation each. The corresponding satellite images had a total of 92403 observations, leading to an overdetermined system (i.e. $92403$ constraints and $627$ degrees of freedom). 

\subsection{Linear least squares with Tikhonov regularization}
We cannot hope to solve the system in Equation~\ref{eqn:linear_system} exactly as it is typically overdetermined\footnote{We have thousands of observations, and only a few hundred degrees of freedom (our a posteriori emission estimates).} and both the simulations and observations have errors. We therefore find a solution vector $x$ using linear least squares with Tikhonov regularization. 

Using (ground/visual) observations of the eruption, we create an a priori estimate of the emission, $x_a$, and incorporate this a priori knowledge into our least squares solver to give preference to solutions close to the a priori. 
We start by replacing our inverted emissions, $x$, with $\tilde{x}=x-x_a$,
\begin{align*}
    Mx &= y_0,\\
    M(x-x_a) &= y_0 - M x_a,\\
    M\tilde{x} &= \tilde{y},
\end{align*}
to penalize solutions that lie far from our a priori. We still cannot find an exact solution to this problem, but we can find the optimal solution, in a least squares sense, by minimizing
\begin{align}
\nonumber
J_1 &= ||M\tilde{x} - \tilde{y}||.
\end{align}
However, the observations are known to have measurement error, and this uncertainty can be included by assigning a weight to each observation
\begin{align}
J_1 &= ||\sigma_o^{-1} \left( M\tilde{x} - \tilde{y} \right)||
\end{align}
in which ${\sigma_o}$ is a diagonal matrix with the standard error of observations to control how close we want our computed solution to match observation ${y_0}_i$. 

In the formulation above, we can control how close we want our solution to lie to the observations. To control how close our solution should lie to the a priori knowledge, we add a second minimization term, 
\begin{align}
J_2 &= ||\sigma_x^{-1} (x - x_a)|| = ||\sigma_x^{-1} \tilde{x}||,
\end{align}
where $\sigma_x$ is a matrix with the estimated standard error of the a priori estimates on the diagonal. In our experiments, we have used a $\sigma_x=\tfrac{1}{2}x_a$.

Unfortunately, solving this minimization problem often results in solutions with sharp gradients, i.e., discontinuous in the vertical dimension. To avoid such unphysical solutions we can add a smoothness minimization term,
\begin{align}
J_3 &= \epsilon ||D\tilde{x}||,
\end{align}
in which $D$ is a diagonal matrix that calculates the second derivative of $\tilde{x}$, and $\epsilon$ determines how smooth we want the solution to be. We have typically used $\epsilon=1.0e-3$, and this parameter is typically set by experimentation.

\begin{figure*}
	\centering
	\begin{subfigure}[b]{0.47\linewidth}%
		\includegraphics[width=\linewidth, trim=3.5cm 3.5cm 7cm 4cm, clip]{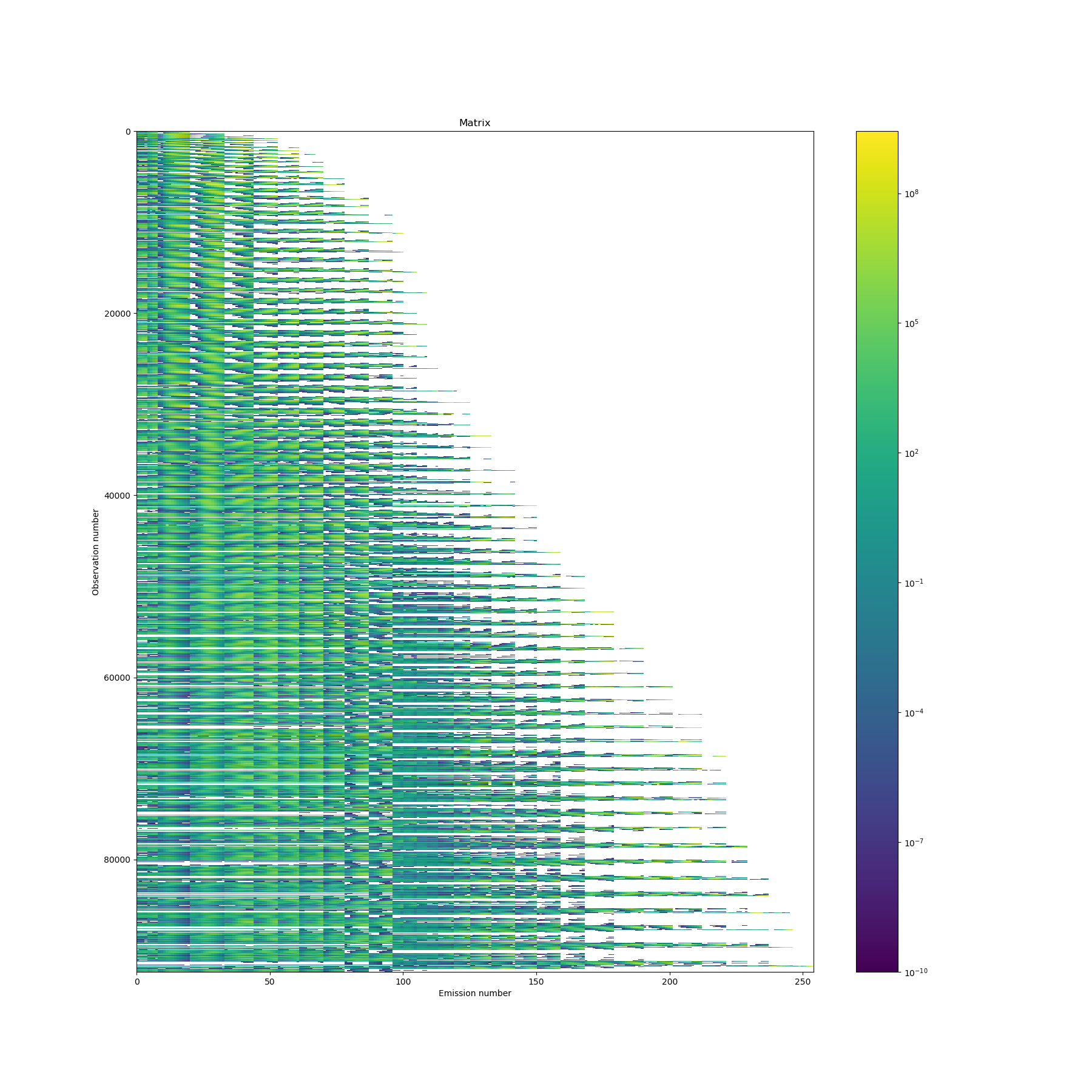}
		\caption{}
		\label{subfig:source_receptor_matrix}
	\end{subfigure}
	\hfill
	\begin{subfigure}[b]{0.47\linewidth}%
		\includegraphics[width=\linewidth, trim=3.5cm 3.5cm 7cm 4cm, clip]{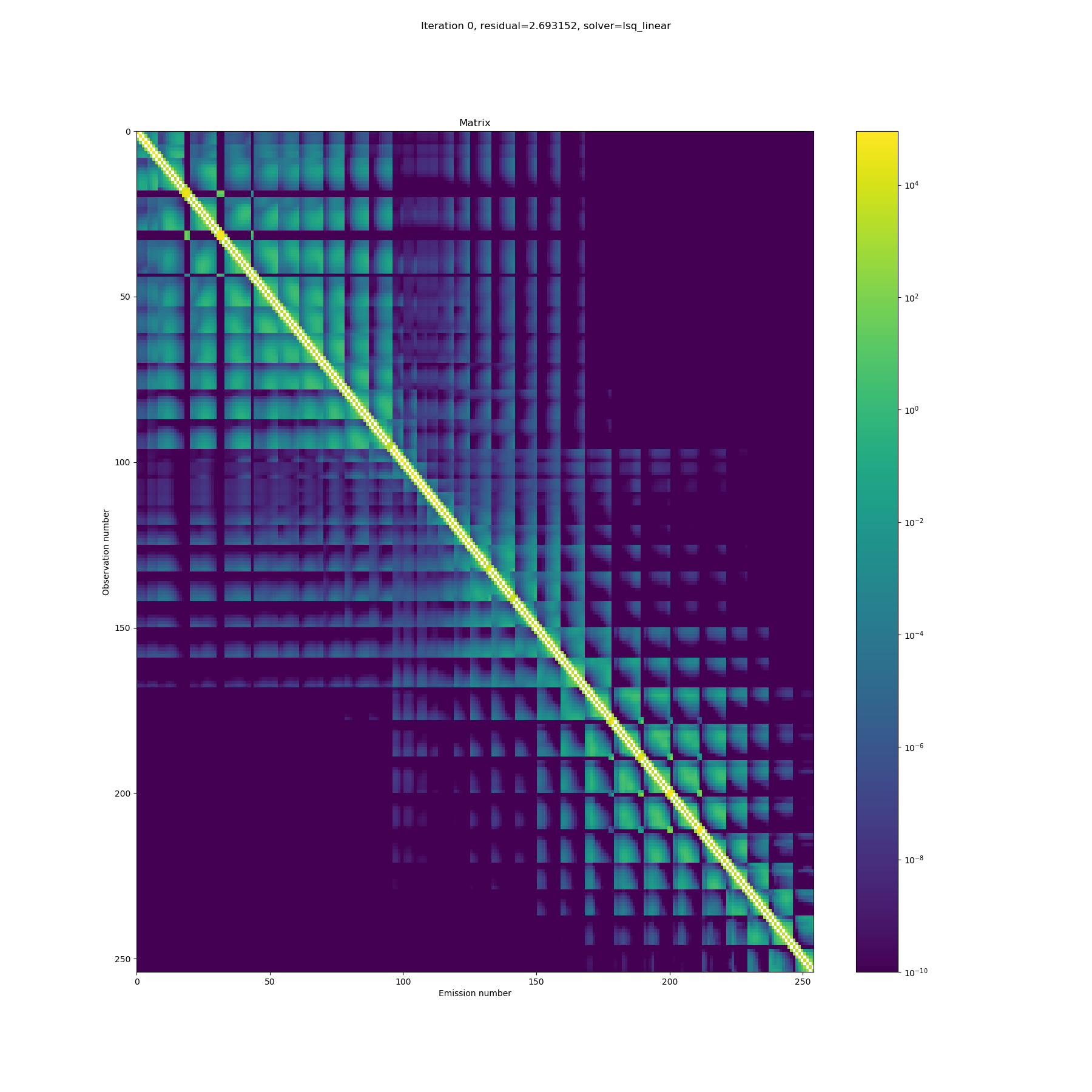}
		\caption{}
		\label{subfig:lsqr_matrix}
	\end{subfigure}
	\caption{Source receptor matrix $M$ (left) and Least squares matrix $G$ (right). The source-receptor matrix for the Eyjafjallajökull case for the period 14-18 April has  92403 rows and 570 columns (after pruning), and the least squares matrix is a square matrix with 570 rows and columns. This then means that the inversion shown uses 92403 observations to estimate the 570 a posteriori emissions.}
	\label{fig:matrices}
\end{figure*}

We can solve these combined minimization problems as a Tikhonov regularization problem (with Tikhonov matrix $Q=\sigma_x^{-2} + \epsilon D^TD$), in which the optimal solution is computed as
\begin{align}
[M^T\sigma_o^{-2}& M
+ \sigma_x^{-2}
+ \epsilon D^TD] \tilde{x}
= M^T \sigma_o^{-2} \tilde{y}\\
x &= x_a + [M^T\sigma_o^{-2} M
+ \sigma_x^{-2}
+ \epsilon D^TD]^{-1}
M^T \sigma_o^{-2} \tilde{y}\\
&= x_a + G^{-1} M^T \sigma_o^{-2} \tilde{y}
\end{align}

Figure~\ref{fig:matrices} shows both the source-receptor matrix $M$ and the least squares matrix $G$ which then is used to compute the a posteriori emission estimates $x$. $M$ is generally a large sparse matrix with $m$ rows (one per observation $O_i$), and $n$ columns (one per emission, $S_j$), whilst $G$ is a relatively small dense matrix with $m$ rows and columns.

It is important that the units of the matrices and vectors are compatible when formulating the minimization problem. Our source-receptor matrix, $M$, and observation vector, $y$, are concentrations per area scaled to $kg/m^2$. The a priori emission estimate, $x_a$ is mass scaled to teragrams ($10^{12}~g$), and our individual simulations $S_j$ have a source term from the volcano that emits one teragram of ash at the given time point $\emist$ and altitude $\emisa$. This means that our computed solution $x$ is given in teragrams of emitted ash.

\subsection{Pruning the linear system of equations}
The system of linear equations outlined above depend on a large amount of input data, and not all data points contribute to the computed solution. By removing parts of the system that have no influence on the solution we can speed up the computational time significantly without reducing the solution quality. We prune based on the a priori knowledge, observations, and simulations.

\paragraph{Pruning a priori values}
We can remove zero a priori values from our system, thereby removing the possibility for our inversion system to set an emission for these altitudes and times. This means that if we are certain that the ash cloud is not higher than a given altitude, or that there have been no emissions for a certain period of time, we can force our a posteriori emission to respect that. We can equivalently allow the inversion procedure to have emissions at these points in time and altitude by using a small non-zero epsilon as a priori value. We also remove all a priori values that do not match up with any simulation or observation, as we have no way of determining these. This reduces the system matrix size from 6061 columns to 589 columns for the Eyjafjallajökull case.

\begin{figure*}
    \centering
    
    \begin{subfigure}[b]{0.33\linewidth}%
    \includegraphics[width=\linewidth]{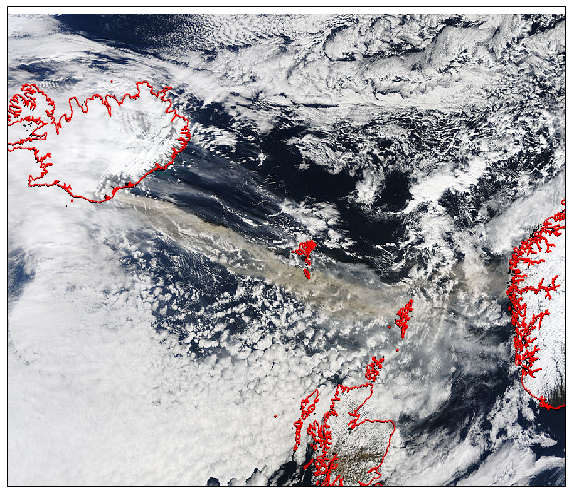} 
    \caption{}
    \label{subfig:satellite_image}
    \end{subfigure}%
    \hfill%
    \begin{subfigure}[b]{0.33\linewidth}%
    \includegraphics[width=\linewidth]{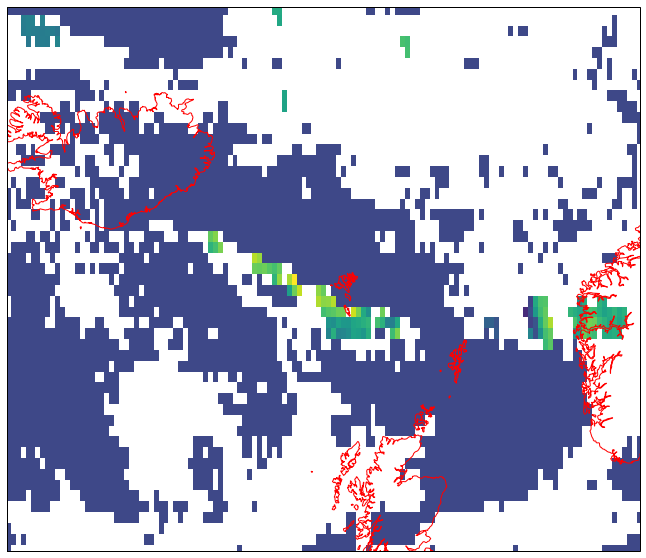} 
    \caption{}
    \label{subfig:satellite_ash_with_zeros}
    \end{subfigure}%
    \hfill%
    \begin{subfigure}[b]{0.33\linewidth}%
    \includegraphics[width=\linewidth]{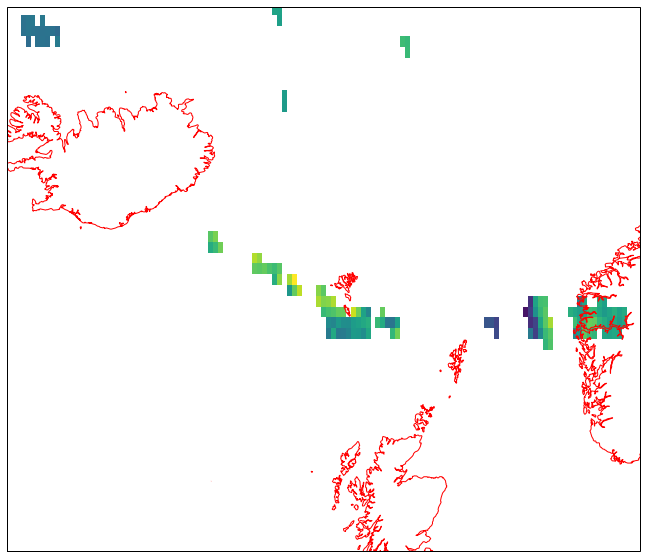}
    \caption{}
    \label{subfig:satellite_ash}
    \end{subfigure}
    \caption{Satellite image and corresponding detection of ash. (\subref{subfig:satellite_image}) is from NASA Terra/MODIS 2010/105 04/15/2010 11:35 UTC, 
    and (\subref{subfig:satellite_ash_with_zeros}) is from the SEVIRI satellite instrument after detection of ash concentrations at 11:00. White pixels are unobserved or uncertain parts of the domain, and blue pixels are observations of zero ash. (\subref{subfig:satellite_ash}) shows only certain non-zero detection of ash.}
    \label{fig:satellite}
\end{figure*}

\paragraph{Pruning observations.}
A satellite image contains both observations of ash and no ash. In addition, cloud cover yields areas where it is difficult to detect ash in satellite images. Figure~\ref{fig:satellite} shows a visible light image compared with the detected ash concentrations. If we use all of these observations, we end up with an enormous linear system that can be even more difficult to solve. We first discard all uncertain observations, as we here cannot determine if there is ash at all. 
We continue by thinning out the zero ash observations, picking at random a fraction of them as in~\cite{stohl2011determination}. This reduces the system matrix size from 18143 rows to 17545 rows for the Eyjafjallajökull case. Note that reducing the number of zero observations here allows the inversion algorithm to place ash in areas where no ash has been observed. This is often a wanted side-effect as a slight difference between true and modeled meteorology may place the ash at a wrong location. Having a high density of zero ash observations in these areas (see e.g., Figure~\ref{subfig:satellite_ash_with_zeros}) will then severely penalize all simulations that have ash in these locations.

\paragraph{Pruning simulations}
Because there is an inherent difference between the actual wind directions and the ones used in our simulation, we end up with an inevitable mismatch between the simulated and observed ash. In addition, there may be false positives of ash in which there is actually no ash. In these cases we will end up with a non-zero observation of ash, but all our simulations have zero ash. This clearly has no contribution to the final system, as all simulations match the observation equally. These zero rows in the matrix (matching non-zero observation) are therefore pruned from the system matrix and reduces the system matrix size from 17545 rows to 15396 rows for the Eyjafjallajökull case. We can also set a threshold so that concentrations beneath a small epsilon are considered zero.

Similarly to the zero pruning based on the row sum, we can remove all zero columns from the matrix. If the sum of a column is zero we know that an emission at the corresponding time and altitude is not visible in any of our observations. Hence, we cannot determine this emission, and remove the column from the matrix and the matching value from our a priori and vector of unknowns. This reduces the system matrix size from 589 columns to 570 columns.

\subsection{Iterative inversion procedure}

Because there are large uncertainties in both our meteorology and satellite observations, this may lead to negative emission estimates at certain points, as there is nothing in our minimization problem that prohibits negative solutions. Negative values in the a posteriori are are forced to lie closer to the a priori estimate by reducing the uncertainties, $\sigma_x$ for these values and recomputing the solution. The iterative procedure repeats until the amount of negative ash emissions is reduced to a fraction (e.g., 1\%) of the total a posteriori emission estimate.

\subsection{Including ash cloud top information}

\begin{figure*}
    \centering
    \begin{subfigure}[b]{0.45\linewidth}%
    \centering%
    %
    \resizebox{!}{6 cm}{%
            \clipbox{0.5cm 3cm 16cm 3cm}{%
                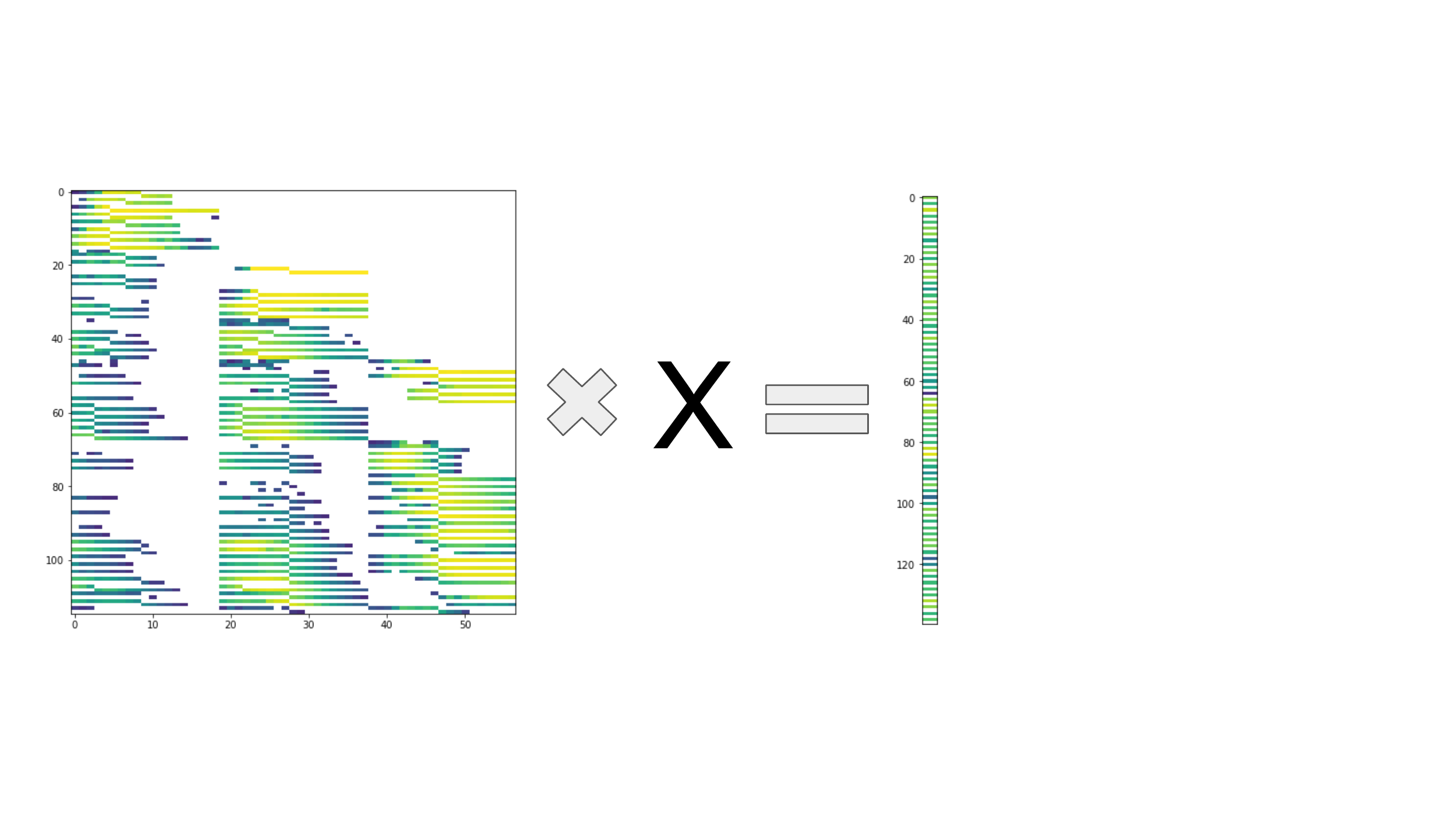%
            }%
    }
    \caption{}%
    \label{fig:source_receptor_matrix_altitude}%
    \end{subfigure}%
    \qquad%
    \begin{subfigure}[b]{0.15\linewidth}%
    \centering%
    %
    \resizebox{!}{6 cm}{%
            \clipbox{15.5cm 3cm 8.5cm 3cm}{%
                \input{svg-inkscape/matrix_obs_alt_svg-tex.pdf_tex}%
            }%
    }
    \caption{}%
    \label{fig:observations_vector_altitude}%
    \end{subfigure}
    \caption{Linear system of equations with altitude information. %
    (\subref{fig:source_receptor_matrix_altitude}) shows the source-receptor matrix $M$ with altitude information, and (\subref{fig:observations_vector_altitude}) the corresponding vector of observations. Compare with Figure~\ref{fig:source_receptor_matrix} and notice that each row is now split into two new rows. Even numbered rows now correspond to observations of ash (below the detected ash plume height), and odd numbered rows correspond to no ash.  
    }
    \label{fig:linear_system_altitude}
\end{figure*}

A novelty in this paper is the use of observed altitude from the SLSTR instrument. The SLSTR instrument can detect the top of the ash plume, and thereby restrict the inversion procedure to give more correct altitudes for the a posteriori emissions. Mathematically, we formulate this by splitting each observation into two observations: one non-zero observation from the ground up to the detected plume height, and one zero observation from the plume height to the top of the model. In essence, we simply split each row in the source-receptor matrix, $M$ into two as shown in Figure~\ref{fig:linear_system_altitude}. This then doubles the number of rows in the matrix, whilst keeping the number of non-zeroes constant. The rest of the algorithm is unchanged.

The computational cost of this extra information is negligible for the whole algorithm as our final linear least squares system, $G$ has the same dimension both with and without altitude information. By representing the system matrix $M$ as a dense matrix, our memory requirement during the algorithm doubles, but is still quite manageable. 

\section{Atmospheric Dispersion Model}
\label{sec:adm}
For this work, we are using the eEMEP model, which is an Eulerian advection model based on the fourth order positive definite advection scheme of Bott~\cite{bott1989positive}. The model is being used operationally to generate volcanic ash forecasts at the Norwegian Meteorological Institute, which are published on the Avinor Internet Pilot Planning Center webpages for all pilots to use. 

\begin{figure}
    \centering
    \includegraphics[width=0.55\linewidth]{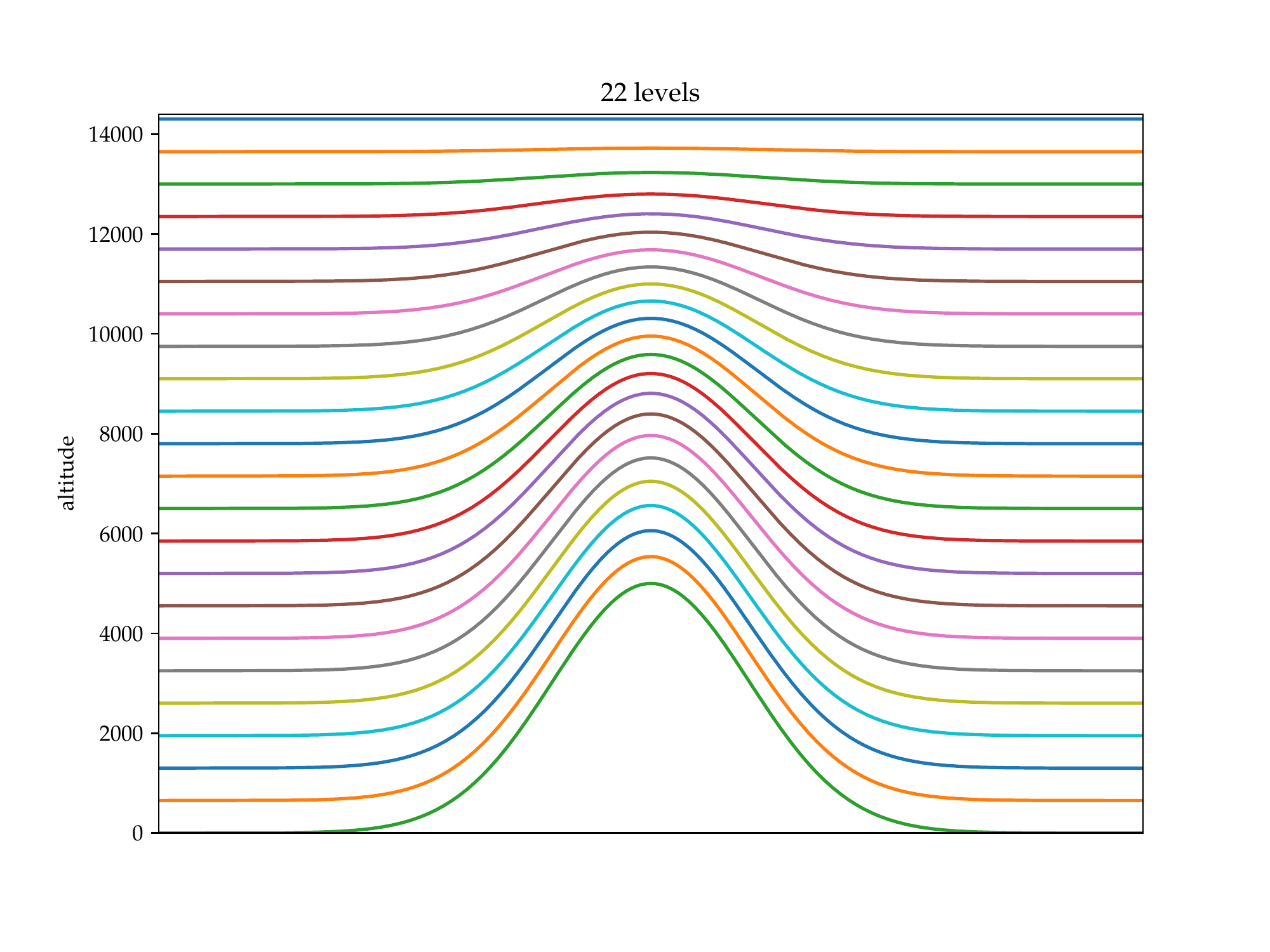}
    \caption{Vertical hybrid sigma levels for the inversion runs. Each level is designed to corresponds to roughly 650 meters of altitude given a ground pressure of 1013.25~hPa. The bottom layer shows a synthetic topography, and how this alters the altitude of the different layers. Please note that these layers are only represented as hybrid sigma coordinates, and never represent actual meters above sea level. The actual vertical definitions file is available in~\cite{zenodo_forward_runs}}
    \label{fig:vertical_levels}
\end{figure}

The current operational setup uses 48 vertical hybrid sigma layers from the ground up to 9.26~hPa (around 26 km above sea level). The layer thickness is smallest close to ground, and increases with altitude. For our use, this may not be the most efficient approach\footnote{The levels closest to ground are typically very thin, which is good from an accuracy point of view for concentrations close to ground. However, when we want to estimate the ash emissions it requires a huge computational effort to handle these layers without them having a significant effect on the result. Most of the volcanic ash we observe is emitted high into the atmosphere.}, and we instead use 22 vertical levels which are close to 650 meters thick each, shown in Figure~\ref{fig:vertical_levels}. This is a trade-off between the number of levels to emit into (which corresponds to the system matrix size), and the computational time required to run the model. We restrict the vertical extent of our model to around 14~km due to limitations in the meteorological input fields available for the 2010 Eyjafjallajökull case.. 

An inversion run with 22 levels of emission every three hours for four days results in over 700 different unit emission scenarios that need to be simulated to generate the source-receptor matrix $M$. A regular run with the simulation model takes around 20 minutes to complete using 32 CPU-cores, which means that this represents over 40 weeks of CPU time. This is prohibitively expensive, and we therefore use a special version of the EMEP model that can run up-to 19 tracers simultaneously. These tracers are independent, and reduces the number of simulator runs from 700 to 36. The major savings come here from only having to process the meteorology 36 times instead of over 700 (which typically is the bottleneck for this kind of application). The numerical advection and writing results to file are not optimized by this approach. For a simpler setup, we also do not use the up-most 3 levels, so that the 19 altitudes at each emission time can be simulated with a single run. We end up with 33 runs that each take around 20 minutes to complete, and using the Nebula supercomputer we are able to get all these inversion runs completed in less than one hour. 

\begin{figure}
    \centering
    \begin{subfigure}[b]{0.99\linewidth}
        \includegraphics[width=\linewidth]{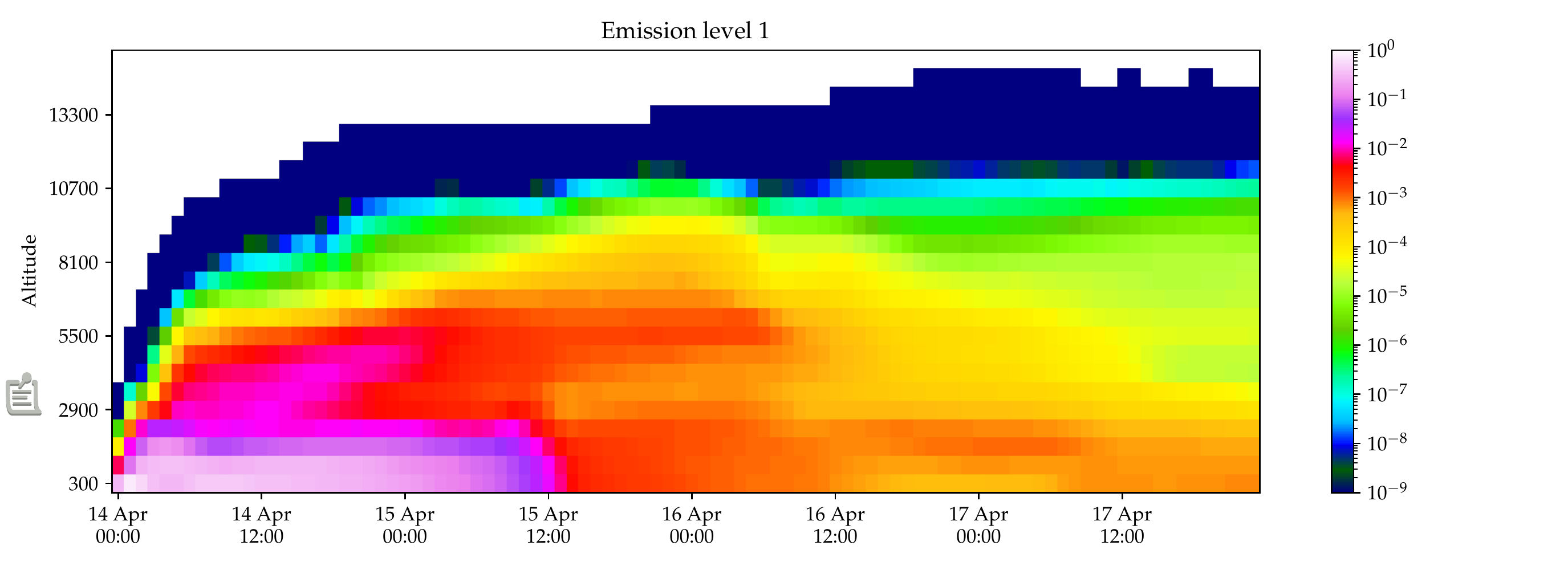}
        \caption{}
        \label{subfig:emis_level_1}
    \end{subfigure}%
    \\%
    \begin{subfigure}[b]{0.99\linewidth}
        \includegraphics[width=\linewidth]{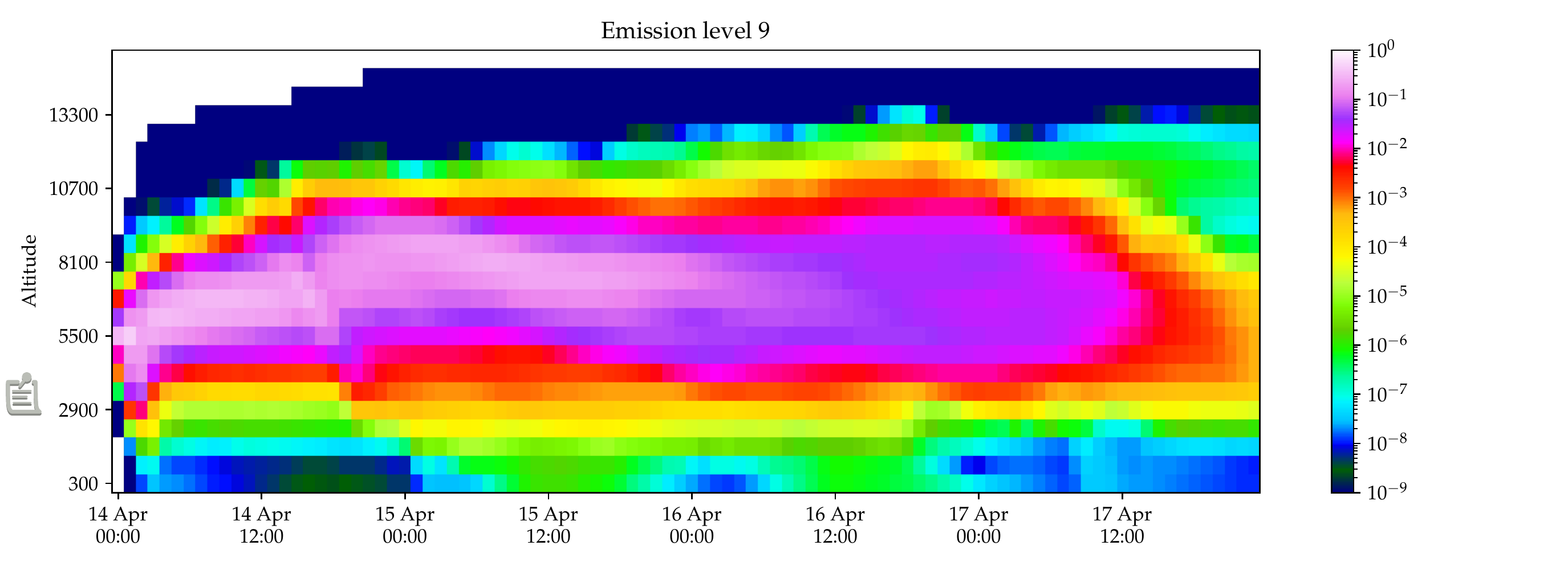}
        \caption{}
        \label{subfig:emis_level_9}
    \end{subfigure}%
    \\%
    \begin{subfigure}[b]{0.99\linewidth}
        \includegraphics[width=\linewidth]{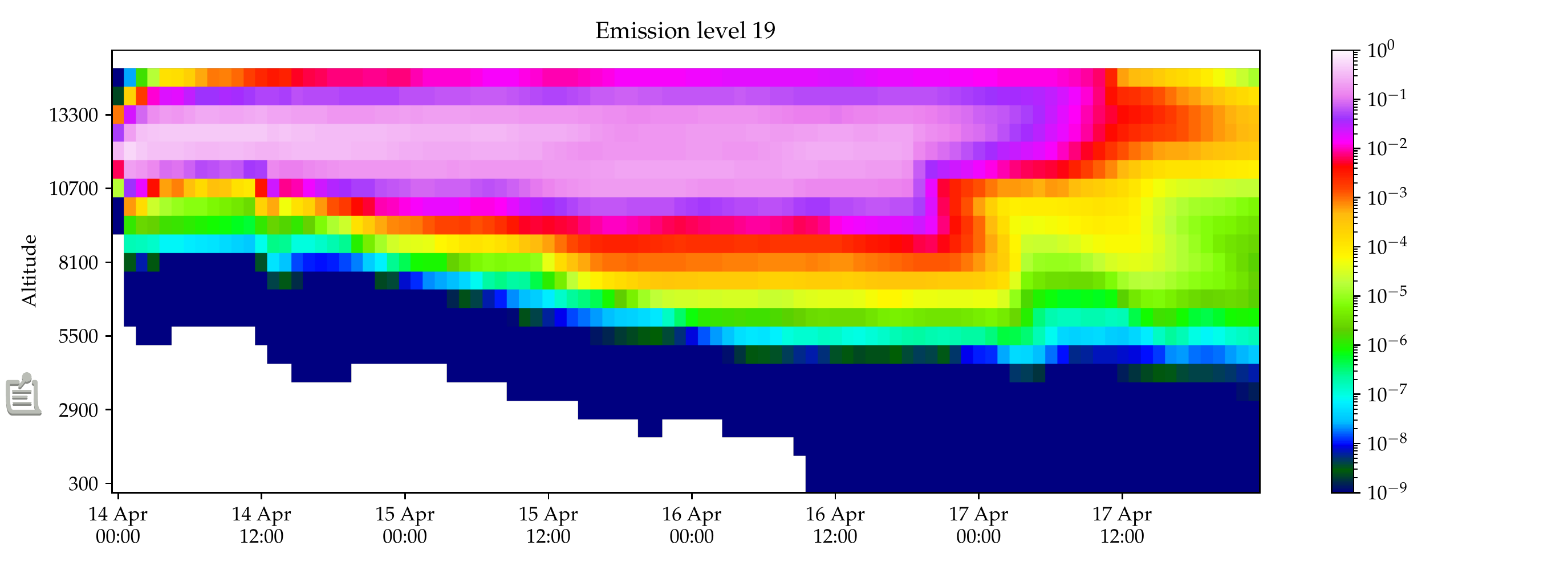}
        \caption{}
        \label{subfig:emis_level_19}
    \end{subfigure}
    \caption{Unit emissions used in the inversion procedure. Here we emit a unit (1 teragram) of ash at different emission levels, and plot the amount of ash in different layers over time. The ash is released at the different levels shown in Figure~\ref{fig:vertical_levels}. Notice that for emissions close to ground, the ash travels only for a short time.}
    \label{fig:emission}
\end{figure}

Figure~\ref{fig:emission} shows the result of emission simulations for different emission altitudes. The three shown simulations emit the ash in level 1, 9, and 19 at midnight on April 14th. The simulation then progresses, and the plot shows how the ash is distributed in the vertical dimension as time goes on. Notice that large parts of the ash cloud leaves the simulation domain just after midday on April 16th. 

\section{Synthetic benchmark cases}
\label{sec:synth}
To check if the inversion procedure presented here works as intended is a non-trivial exercise, as there are large uncertainties in both the simulation model and observations being used. 
We have therefore checked our inversion procedure against a known truth, generated by the simulation model itself. We first generate an a priori emission estimate, and use this a priori estimate to generate the synthetic truth consisting of satellite images. These synthetic satellite images observe the synthetic truth at random locations in space and time. We then expect that our inversion procedure should generate an a posteriori which lies close to the a priori used to generate the truth. It should be noted that we do not expect a perfect inversion, as we try to estimate the vertical and time distribution of the ash emission whilst we only observe the vertically integrated ash concentration at certain points. 
We have used the Eyjafjallajökull 2010 eruption as a basis to generate a realistic scenario from April 14th to 18th. We also vary the a priori estimate and the ash cloud top altitude to see the effect of these parameters on the solution. 
 It should also be noted that synthetic benchmarks removes any uncertainty in both the transport model and the numerical weather forecast data.

\subsection{Varying a priori}
\begin{figure*}
    \centering
    \begin{subfigure}[b]{\linewidth}%
        \includegraphics[width=\linewidth]{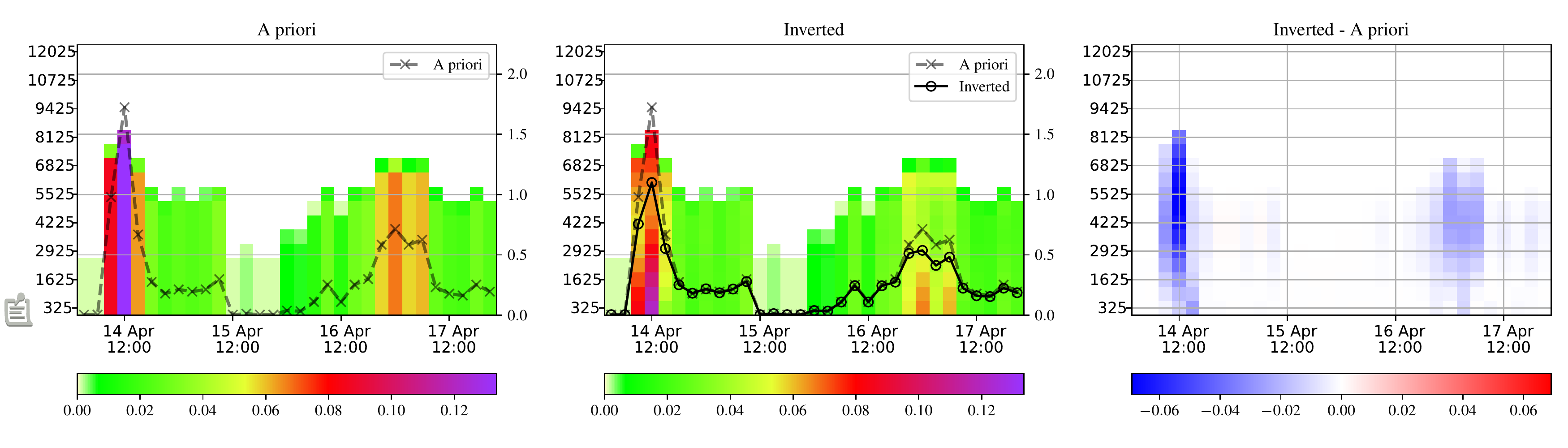}
        \caption{}
        \label{subfig:eyja_synthetic_double}
    \end{subfigure}
    \begin{subfigure}[b]{\linewidth}%
        \includegraphics[width=\linewidth]{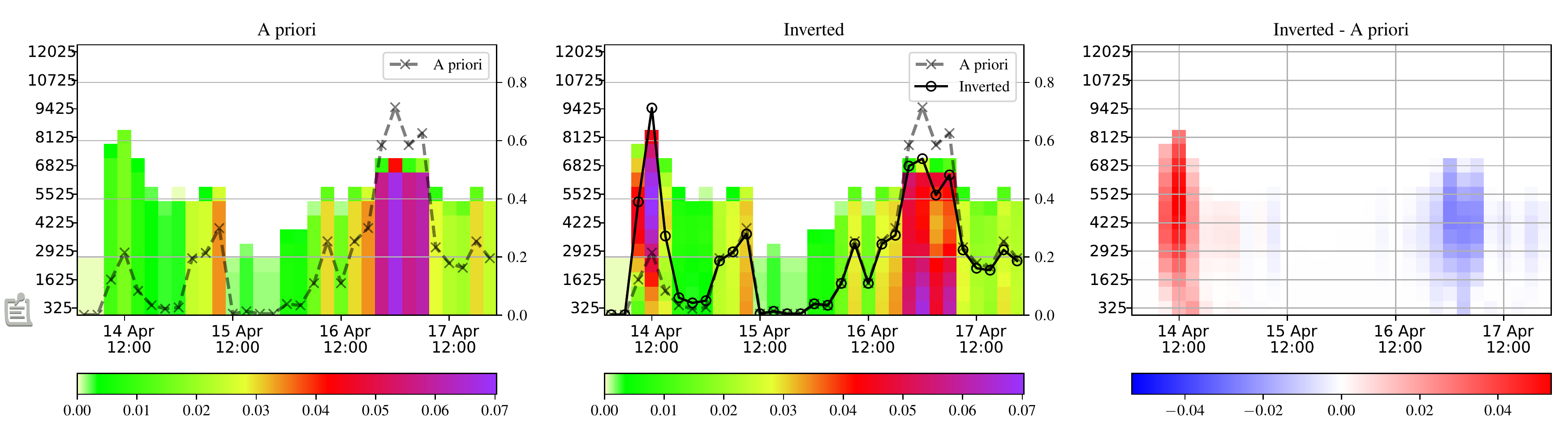}
        \caption{}
        \label{subfig:eyja_synthetic_quarter_then_double}
    \end{subfigure}
    \begin{subfigure}[b]{\linewidth}%
        \includegraphics[width=\linewidth]{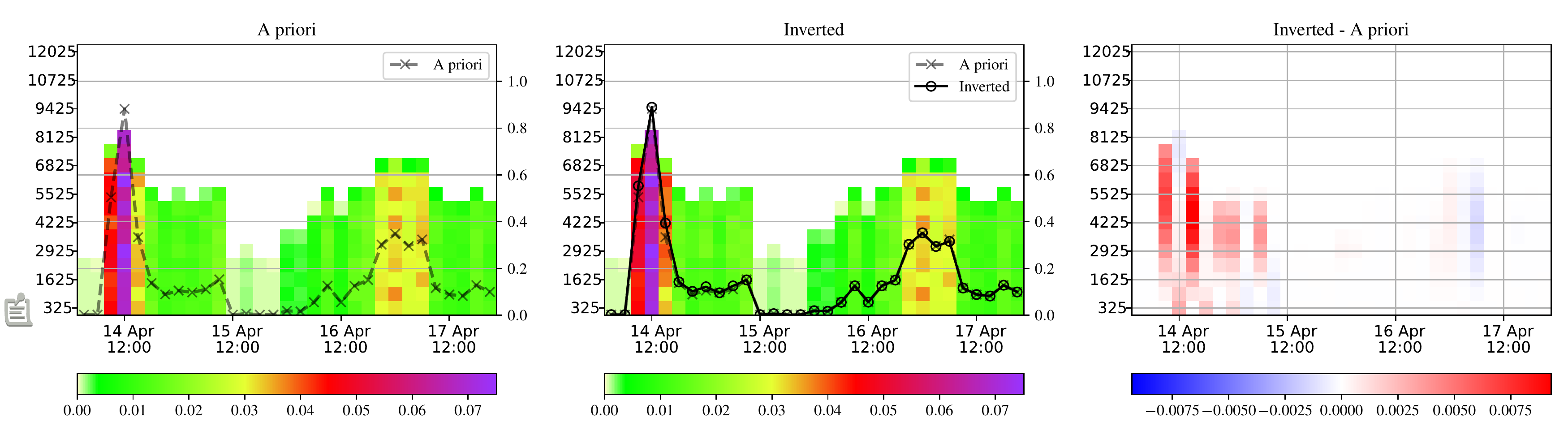}
        \caption{}
        \label{subfig:eyja_synthetic_random}
    \end{subfigure}
    \caption{Effect of varying the a priori ash emission estimate. The plot shows the sum of the a priori and a posteriori emissions for each time-point (in teragrams), and the colors show the emissions at different times and altitudes (also in teragrams).
    The a priori is used to generate a synthetic truth, and with no perturbation the a posteriori is identical (within the expected numerical error) to the a priori. In (\subref{subfig:eyja_synthetic_double}) the a priori is doubled, and the a posteriori shows a clear decrease. In (\subref{subfig:eyja_synthetic_quarter_then_double}) the a priori is $\tfrac{1}{4}$ for the first half of the period, and doubled for the last. In (\subref{subfig:eyja_synthetic_random}) we add a 25~\% random noise to the a priori.}
    \label{fig:synthetic}
\end{figure*}

The first thing we test with our inversion algorithm is how it behaves when we vary the a priori whilst keeping the observations and the rest of the algorithm fixed. 
We first generate a synthetic satellite image from an a priori emission estimate. We then change the a priori, and compare the a posteriori emission with the known truth. By varying the a priori in this way, we see how sensitive the algorithm is to this important parameter, and Figure~\ref{fig:synthetic} shows results of these experiments. This sensitivity study also uncovers where the inversion procedure has sufficient information to alter the a priori estimate. By examining the results, we see that the a posteriori differs most for the major eruption time points.

\subsection{Synthetic ash cloud top observation}

\begin{figure*}
    \centering
    \begin{subfigure}[b]{0.95\linewidth}
    \includegraphics[width=\linewidth]{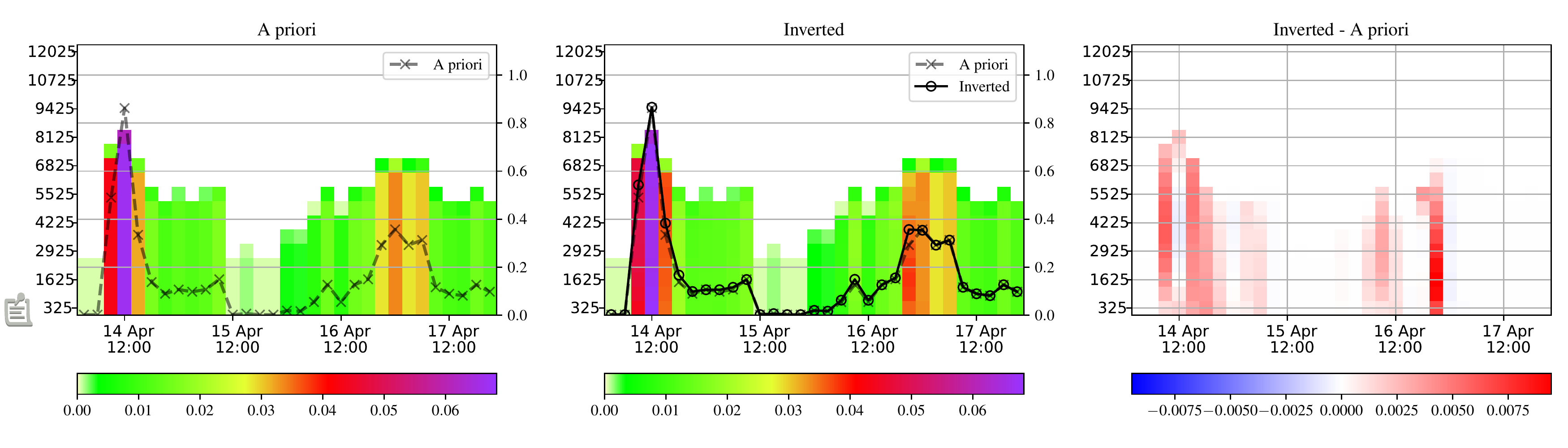}
    \caption{}
    \label{subfig:normal_altitude}
    \end{subfigure}%
    \\%
    \begin{subfigure}[b]{0.95\linewidth}
    \includegraphics[width=\linewidth]{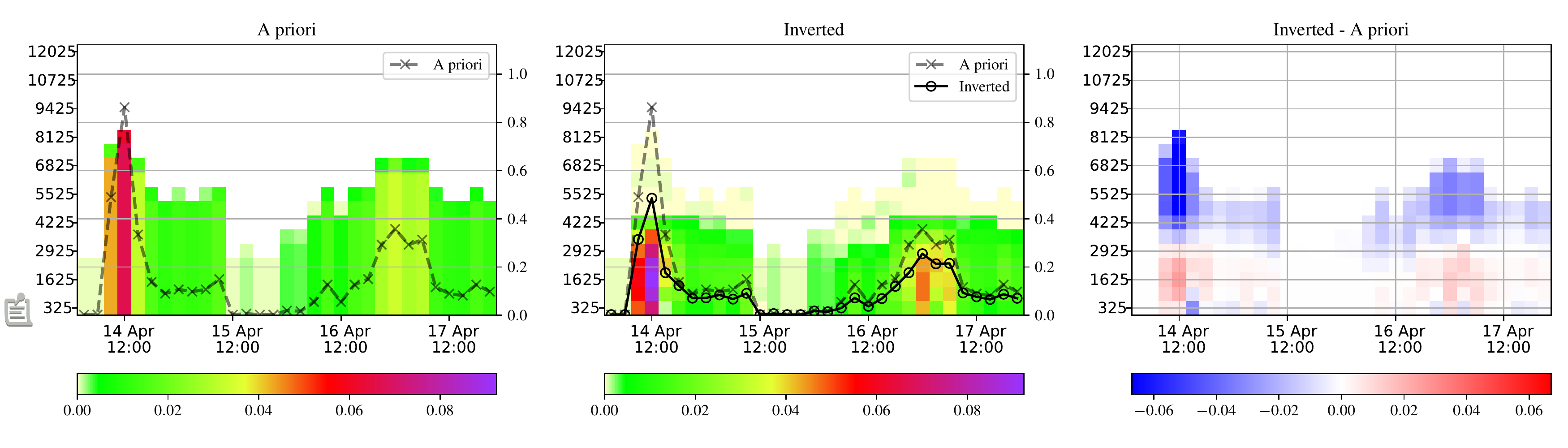}
    \caption{}
    \label{subfig:reduced_altitude}
    \end{subfigure}
    \caption{Effect of using a synthetic altitude observation. In (\subref{subfig:normal_altitude}) we have used the true synthetic ash cloud top in the inversion, whilst in (\subref{subfig:reduced_altitude}) we have reduced the synthetic ash cloud observation to half the altitude. }
    \label{fig:altitude}
\end{figure*}

We have also used a synthetic truth to check how our algorithm performs with altitude information in the inversion procedure.
We have generated a synthetic satellite image with height information directly from the simulation results. This means that we should be able to get an inverted a posteriori emission estimate which is very close to our a priori. We have run two experiments, one with the true top of the ash cloud, and one in which we simply reduce this altitude to half of the true top height (here, the only difference between the two datasets is in the altitude information - the ash mass per square meter is identical).

Figure~\ref{fig:altitude} shows the effect of using altitude in the inversion procedure. The figure shows how using the true top altitude of the synthetic ash cloud does not change the inversion results, whilst if we halve this height the inverted result shows almost no ash emitted in these high altitudes. Notice that the inversion algorithm is very good at reducing the emissions in the top levels, but does not increase the emissions sufficiently in the lower levels to fully compensate for the missing ash mass. This is a similar behavior as we see when we change the a priori estimate.

\section{Real-world cases}
\label{sec:realworld}
The previous section outlined some of the sensitivities of the inversion procedure on synthetic datasets, in which we have generated satellite images from a synthetic known truth. In this section, we apply the methodology to two real-world eruptions, the Eyjafjallajökull 2010 eruption, and the 2018 Shiveluch eruption. The Eyjafjallajökull eruption is a well studied case which makes it possible to compare the quality of our results with other approaches, and the Shiveluch 2018 eruption was captured by the SLSTR instrument with ash cloud top observations.

\subsection{Eyjafjallajökull 2010 eruption}
\label{sec:eyja2010}
\begin{figure*}
    \centering
    \begin{subfigure}[b]{\linewidth}
        \includegraphics[width=\linewidth]{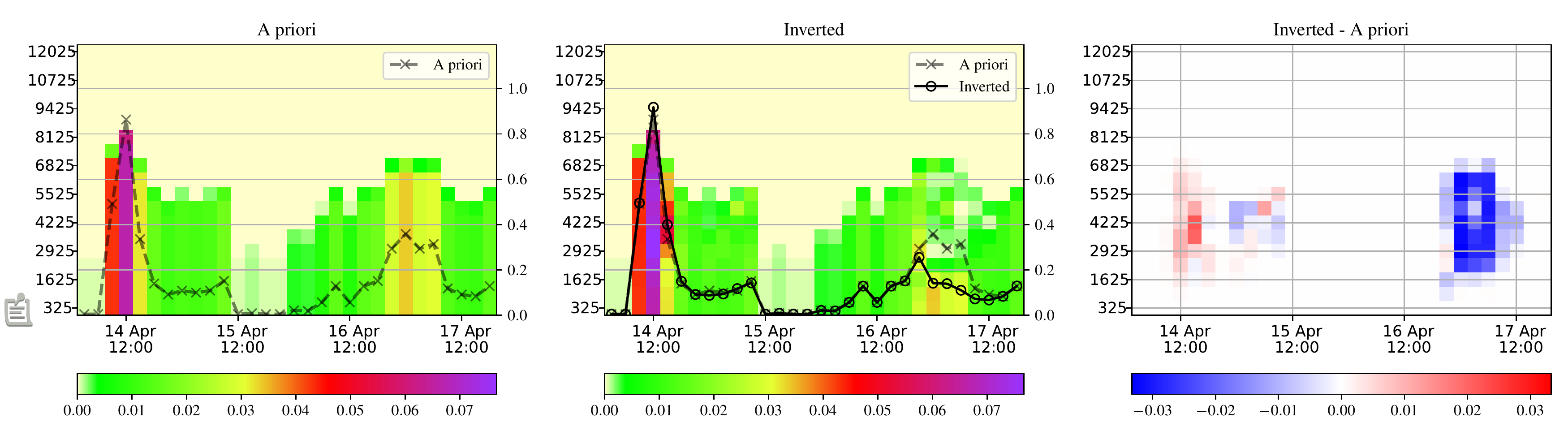}
        \caption{}
        \label{subfig:eyja_real}
    \end{subfigure}
    \begin{subfigure}[c]{\linewidth}
        \includegraphics[width=0.45\linewidth, trim=0 -0.5cm 0cm -0.5cm, clip]{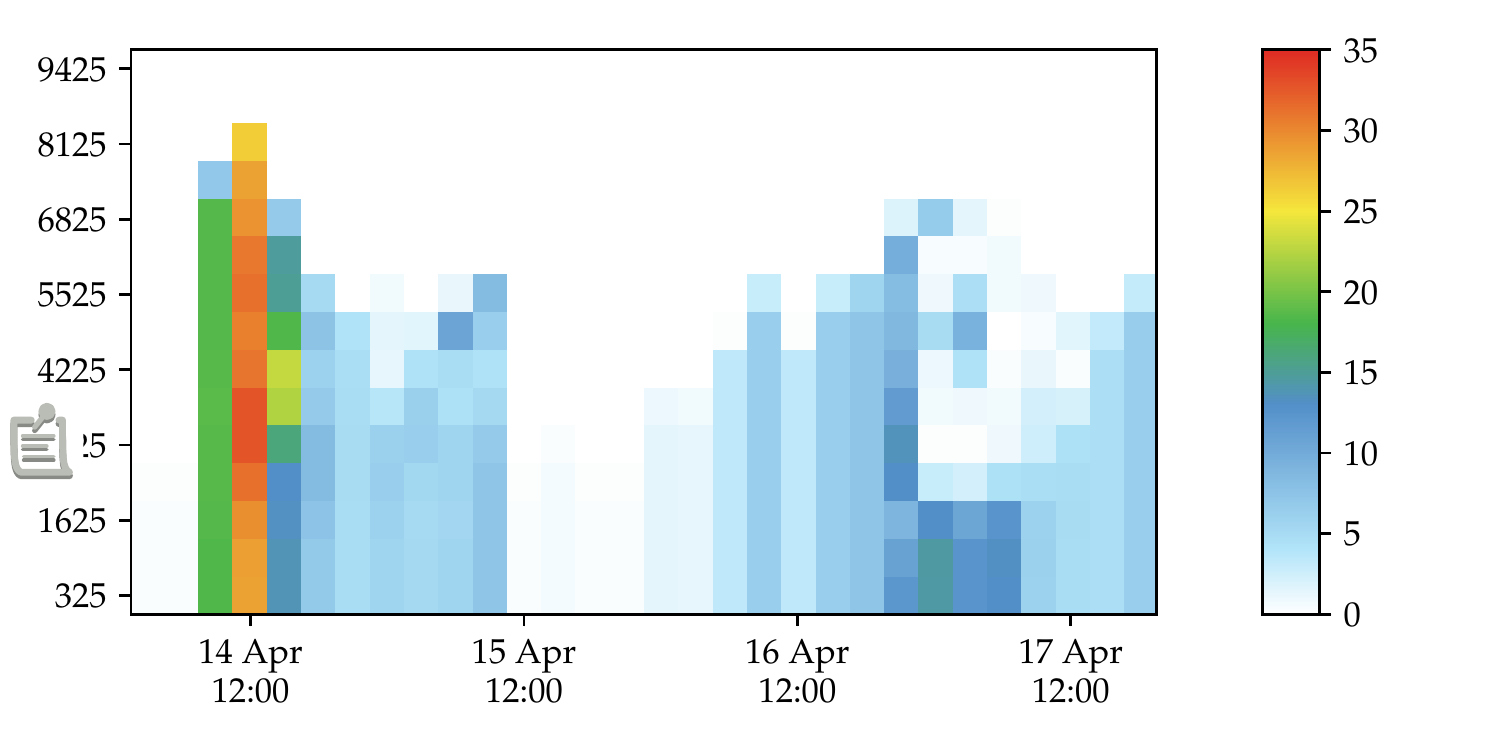}%
        \qquad%
        \includegraphics[width=0.47\linewidth, trim=1.3cm 0 0 1cm, clip]{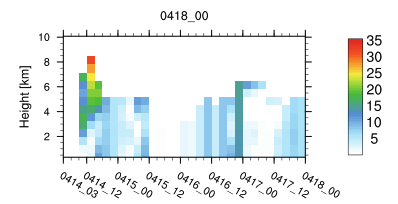}
        \caption{}
        \label{subfig:eyja_compare_birthe}
    \end{subfigure}
    \caption{Eyjafjalla a priori vs inverted emission. (\subref{subfig:eyja_real}) compares the a priori estimate with the inverted results (in teragrams), and the difference shown in the right-most figure. (\subref{subfig:eyja_compare_birthe}) compares our current results with those published by Steensen et~al.~\cite[Figure 6e]{steensen2017uncertainty} (in $kg/(m\cdot s)$).}
    \label{fig:eyja_real}
\end{figure*}

SEVIRI satellite data where taken from \citet{steensen2017uncertainty}. In figure~\ref{fig:eyja_real} is shown the result of the inversion algorithm applied to dataset. Overall the two inversions produce similar results. Some differences are noted: 1) we find that the main emissions for 14 April are fairly evenly spread over the entire column and with a maximum around 3~km, while in \citet{steensen2017uncertainty} the emissions peaks around 8~km and there is no emission below 3~km; 2) early on 17 April we find emissions below 1.6~km while none are reported by \citet{steensen2017uncertainty}. It is noted that results presented here agree better with those presented by \citep[][their Fig.~2c]{stohl2011determination}. The reasons for the differences with \citet{steensen2017uncertainty}: this is a new implementation and there is the possibility for bugs in either the old or new code; there are different parameters used in the simulation runs\footnote{e.g., use of gravitational settling}; there are different parameters used in the inversion run; and the advection model (eEMEP) has gone through significant upgrades and changes. As the details required to reproduce the runs of \citet{steensen2017uncertainty} are not fully available, it is not possible to reproduce their results.

As in the previously published results, the algorithm changes the a priori knowledge in the places in which there is sufficient information, and keeps the a priori estimate for otherwise. Also, it appears more likely to reduce the emission estimate than to increase it. 

When we look at the two results and compare them qualitatively, it appears that the current version places the maximum emission height lower than previous results. The current version also appears to produce more continuous emissions in the vertical (i.e., whereas the old version has zero emissions close to ground the current has emissions in the whole column).

\begin{figure}
    \centering
    \begin{subfigure}[b]{0.65\linewidth}
    \includegraphics[width=\linewidth, trim=5cm 1cm 30cm 1cm, clip]{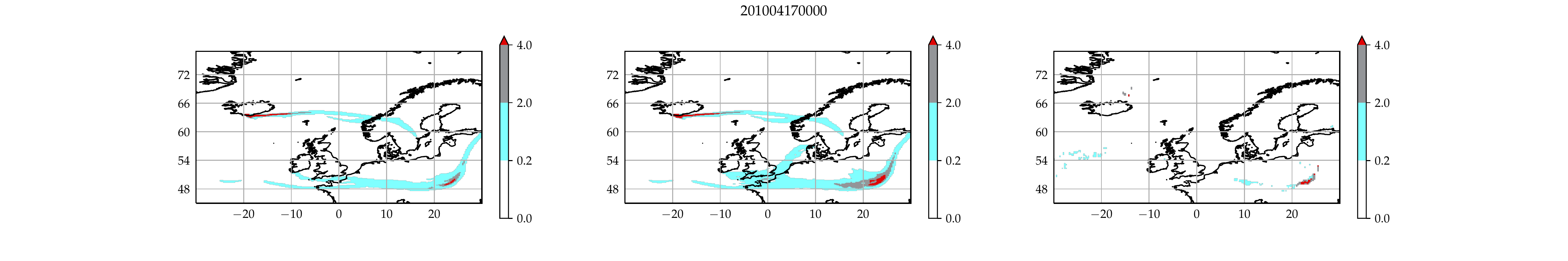}
        \caption{}
        \label{subfig:eyja_apriori}
    \end{subfigure}\\
    \begin{subfigure}[b]{0.65\linewidth}
    \includegraphics[width=\linewidth, trim=17.5cm 1cm 17.5cm 1cm, clip]{figures/eyja_2010/result_201004170000}
        \caption{}
        \label{subfig:eyja_aposteriori}
    \end{subfigure}\\
    \begin{subfigure}[b]{0.65\linewidth}
    \includegraphics[width=\linewidth, trim=30cm 1cm 5cm 1cm, clip]{figures/eyja_2010/result_201004170000}
        \caption{}
        \label{subfig:eyja_satellite}
    \end{subfigure}
    \caption{Ash-concentrations over Europe during the Eyjafjallajökull 2010 eruption on 2010-04-17 00:00Z with 
    (\subref{subfig:eyja_apriori}) a priori, 
    (\subref{subfig:eyja_aposteriori}) a posteriori, and
    (\subref{subfig:eyja_satellite}) satellite image. 
    The ash concentrations are shown in grams per square meter, and classified into the groups no ash ($<0.2~g/m^2$), low ash ($0.2~g/m^2 - 2.0~g/m^2$), medium ash ($2.0g/m^2 - 4.0~g/m^2$), and high ash ($>4.0~g/m^2$) ash concentrations (the same categories used in volcanic ash advisories).}
    \label{fig:eyja_aposteriori}
\end{figure}

Figure~\ref{fig:eyja_aposteriori} shows the result of the inversion procedure compared with the satellite data for a single timestep. By redistributing the ash in the vertical column as shown in Figure~\ref{fig:eyja_real}, the inversion procedure is able to create a much higher ash concentration in the lower right hand corner. This matches well with the observed ash at the same location.

\subsection{Shiveluch 2018 eruption}
Shiveluch is a volcano on the Kamchatka peninsula in the Russian far east which erupts regularly, and the January 2018 eruption was captured by the SLSTR instrument. This gives real-world data that can be used to examine the effect of adding ash cloud height information to our linear system of equations. Unfortunately, the satellite was not able to capture sufficient information for a meaningful inversion\footnote{The amount of observations required to perform a meaningful inversion depends on many factors. We should preferably observe larger parts of an ash plume over time.}. We have nevertheless run the inversion on this limited data set to examine if there are any adverse side effects. 

We have generated an a priori emission estimate based on the volcanic ash advisory messages that indicate the ash plume height at different time points. Our a priori estimate is based on the following information:
\begin{itemize}
    \item Summit/Vent: 3283 m ASL
    \item 20180109 22:43: Ash observed in FL170 (plume at 5200~m ASL)
    \item 20180110 11:20: Ash observed in FL360 (plume at 10950~m ASL)
    \item 20180111 00:48: Residual ash observed in FL310 (eruption ceased)
\end{itemize}

\begin{figure*}
    \centering
    \begin{subfigure}[b]{0.55\linewidth}
        \includegraphics[width=\linewidth, trim=7cm 0cm 3cm 1cm, clip]{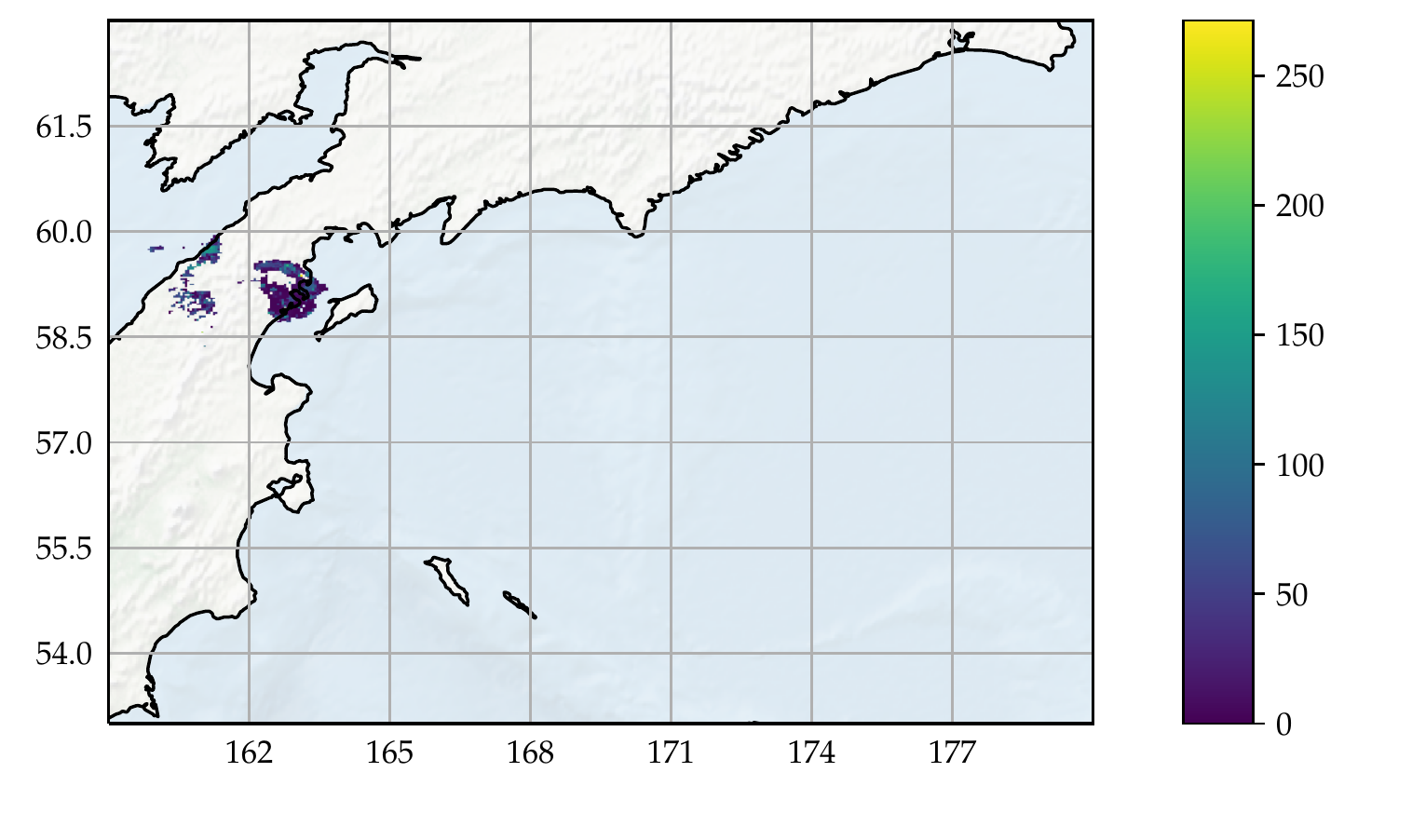}%
        \caption{}
        \label{subfig:shiveluch_satellite}
    \end{subfigure}%
    \\
    \begin{subfigure}[b]{\linewidth}
        \includegraphics[width=\linewidth]{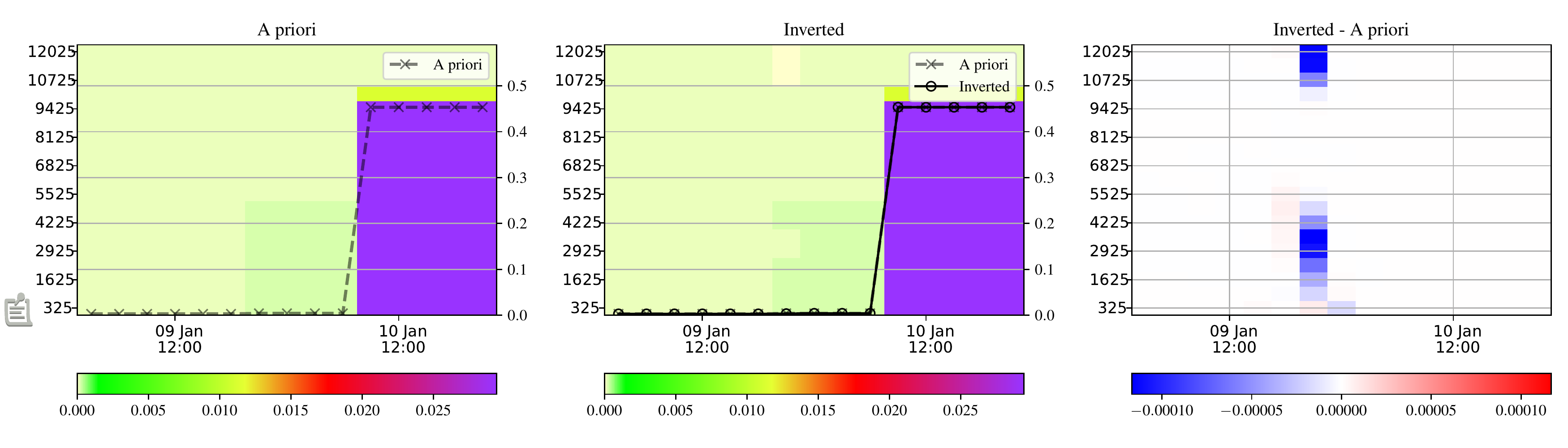}
        \caption{}
        \label{subfig:shiveluch_a_posteriori}
    \end{subfigure}
    \caption{Shiveluch January 2018 eruption. The top image shows the satellite image available for the inversion. Notice that there is very little ash cloud information visible. The bottom figure shows the result of the inversion procedure, which for all practical purposes does not change the a priori estimate.}
    \label{fig:shiveluch}
\end{figure*}

As Figure~\ref{fig:shiveluch} shows, there appears to be no negative side effects of running the inversion with insufficient data, but the value of the inversion is also negligible.

\section{Summary}
\label{sec:summary}
We have presented an inversion algorithm for volcanic ash emission estimates based on satellite imagery, and the main novelty in this work lies in the inclusion of ash cloud top altitude information and pruning of the system of linear equations. We show that by splitting an observation into two observations (one of ash up to the ash cloud top, and one of no ash above the ash cloud top), we are able to constrain the solution to emit in the correct altitudes. By pruning the system of linear equations, we dramatically reduce the total system size and decrease time to solution. Our results are different from previously presented results, but our inverted results match well with satellite observations. We have also run the inversion procedure on the 2018 Shiveluch eruption, but the limited number of observations means that the inversion also adds limited value. 

All source code used in this work is released under an open source license, and available on Github~\citep{zenodo_software}. The datasets used are also available under open licenses~\citep{zenodo_forward_runs,zenodo_satellite}.

\subsection{Future work}
This work has several opportunities for further improvement that we see natural to pursue. First of all, the inversion runs have been run with a version of eEMEP which does not properly include the gravitational settling effects. Upgrading the handling of these effects will most probably increase the performance of the inversion algorithm as the numerical model will be closer to the physical model. 

This work has also uncovered some of the sensitivities of the inversion procedure. It will be very interesting and important to perform an uncertainty quantification of the inversion algorithm, in order to pinpoint which a posteriori emissions are certain, and which are uncertain. In conjunction with this, it will be very interesting to run the inversion algorithm on other synthetic truth data. One of the major sources for uncertainties in these kinds of inversions is the meteorology, and we aim to create a synthetic truth using FLEXPART and NCAR GFS meteorology, and run the inversion with eEMEP and ECMWF IFS meteorology. That way we can generate a synthetic truth (with altitude information) that we know we are not able to fully reproduce with our inversion. 

The current approach of ash inversion is highly sensitive to the meteorology, and a small difference between the true and modeled meteorology may create a posteriori emissions that are far from the true emissions. It will be highly interesting to rephrase the inversion algorithm into a basis which is less sensitive to the meteorology, e.g., using streamline distance from the volcano or similar. Also using other approaches than least squares to solve the minimization problem will be an interesting line of pursuit.

\subsection{CRediT statement}
A. R. Brodtkorb: Methodology Development, Validation, Data Curation, Writing - Original Draft, Writing - Review \& Editing, Visualization;
A. Benedictow: Methodology	Development;
A. Kylling: Conceptualization, Writing - Review \& Editing, Funding acquisition;
H. Klein: Conceptualization, Writing - Review \& Editing, Funding acquisition;
A. Nyiri: Data Curation;
A. Valdebenito: Methodology	Development;

\subsection{Acknowledgments}
This work has partly been performed as part of project number NIT.09.16.05 funded by the Norwegian Space Agency. Simulations have been run on the research and development Nebula supercomputer funded by the MetCoOp HPC infrastructure. 

\bibliographystyle{unsrt}
\bibliography{main.bbl}

\end{document}

%% file: svg-inkscape/observations_svg-tex.pdf_tex
\begingroup%
  \makeatletter%
  \providecommand\color[2][]{%
    \errmessage{(Inkscape) Color is used for the text in Inkscape, but the package 'color.sty' is not loaded}%
    \renewcommand\color[2][]{}%
  }%
  \providecommand\transparent[1]{%
    \errmessage{(Inkscape) Transparency is used (non-zero) for the text in Inkscape, but the package 'transparent.sty' is not loaded}%
    \renewcommand\transparent[1]{}%
  }%
  \providecommand\rotatebox[2]{#2}%
  \newcommand*\fsize{\dimexpr\f@size pt\relax}%
  \newcommand*\lineheight[1]{\fontsize{\fsize}{#1\fsize}\selectfont}%
  \ifx\svgwidth\undefined%
    \setlength{\unitlength}{720bp}%
    \ifx\svgscale\undefined%
      \relax%
    \else%
      \setlength{\unitlength}{\unitlength * \real{\svgscale}}%
    \fi%
  \else%
    \setlength{\unitlength}{\svgwidth}%
  \fi%
  \global\let\svgwidth\undefined%
  \global\let\svgscale\undefined%
  \makeatother%
  \begin{picture}(1,0.5625)%
    \lineheight{1}%
    \setlength\tabcolsep{0pt}%
    \put(0,0){\includegraphics[width=\unitlength,page=1]{observations_svg-tex.pdf}}%
  \end{picture}%
\endgroup%

%% file: svg-inkscape/matrix_svg-tex.pdf_tex
\begingroup%
  \makeatletter%
  \providecommand\color[2][]{%
    \errmessage{(Inkscape) Color is used for the text in Inkscape, but the package 'color.sty' is not loaded}%
    \renewcommand\color[2][]{}%
  }%
  \providecommand\transparent[1]{%
    \errmessage{(Inkscape) Transparency is used (non-zero) for the text in Inkscape, but the package 'transparent.sty' is not loaded}%
    \renewcommand\transparent[1]{}%
  }%
  \providecommand\rotatebox[2]{#2}%
  \newcommand*\fsize{\dimexpr\f@size pt\relax}%
  \newcommand*\lineheight[1]{\fontsize{\fsize}{#1\fsize}\selectfont}%
  \ifx\svgwidth\undefined%
    \setlength{\unitlength}{720bp}%
    \ifx\svgscale\undefined%
      \relax%
    \else%
      \setlength{\unitlength}{\unitlength * \real{\svgscale}}%
    \fi%
  \else%
    \setlength{\unitlength}{\svgwidth}%
  \fi%
  \global\let\svgwidth\undefined%
  \global\let\svgscale\undefined%
  \makeatother%
  \begin{picture}(1,0.5625)%
    \lineheight{1}%
    \setlength\tabcolsep{0pt}%
    \put(0,0){\includegraphics[width=\unitlength,page=1]{matrix_svg-tex.pdf}}%
  \end{picture}%
\endgroup%

%% file: svg-inkscape/matrix_obs_alt_svg-tex.pdf_tex
\begingroup%
  \makeatletter%
  \providecommand\color[2][]{%
    \errmessage{(Inkscape) Color is used for the text in Inkscape, but the package 'color.sty' is not loaded}%
    \renewcommand\color[2][]{}%
  }%
  \providecommand\transparent[1]{%
    \errmessage{(Inkscape) Transparency is used (non-zero) for the text in Inkscape, but the package 'transparent.sty' is not loaded}%
    \renewcommand\transparent[1]{}%
  }%
  \providecommand\rotatebox[2]{#2}%
  \newcommand*\fsize{\dimexpr\f@size pt\relax}%
  \newcommand*\lineheight[1]{\fontsize{\fsize}{#1\fsize}\selectfont}%
  \ifx\svgwidth\undefined%
    \setlength{\unitlength}{720bp}%
    \ifx\svgscale\undefined%
      \relax%
    \else%
      \setlength{\unitlength}{\unitlength * \real{\svgscale}}%
    \fi%
  \else%
    \setlength{\unitlength}{\svgwidth}%
  \fi%
  \global\let\svgwidth\undefined%
  \global\let\svgscale\undefined%
  \makeatother%
  \begin{picture}(1,0.5625)%
    \lineheight{1}%
    \setlength\tabcolsep{0pt}%
    \put(0,0){\includegraphics[width=\unitlength,page=1]{matrix_obs_alt_svg-tex.pdf}}%
  \end{picture}%
\endgroup%